\documentstyle[11pt,epsfig,oldlfont,rotating]{article}
\newcommand{\mr}[1]{{\mathrm{#1}}}
\newcommand{\ds}{\displaystyle}

\newcommand{\bbar}[1]{$\overline{\rm #1}$}
\newcommand{\bbare}[1]{\overline{\rm #1}}

\newcommand{\as}{$\alpha_s\;$}
\newcommand{\ase}{\alpha_s}

\newcommand{\reff}[1]{(\ref{#1})}

\newcommand{\be}{\begin{equation}}
\newcommand{\ee}{\end{equation}}
\newcommand{\bd}{\begin{description}}
\newcommand{\ed}{\end{description}}
\newcommand{\bmat}{\begin{displaymath}}
\newcommand{\emat}{\end{displaymath}}
\newcommand{\bit}{\begin{itemize}}
\newcommand{\eit}{\end{itemize}}
\newcommand{\ben}{\begin{enumerate}}
\newcommand{\een}{\end{enumerate}}
\begin{document}
\begin{flushright}
PRA--HEP/95--1
\end{flushright}
\begin{center}
{\Large \bf On the role of the quark mass thresholds in
extrapolations of the running \as} \\
\vspace*{0.3cm}
{\em Ji\v{r}\'{\i} Ch\'{y}la} \\
\vspace*{0.3cm}
Institute of Physics, Academy of Sciences of the Czech Republic  \\
Na Slovance 2, Prague 8, 18040 Czech Republic \\
\vspace*{0.8cm}
{\bf Abstract}   \\
\end{center}

\noindent
The accuracy of the conventional treatment of quark mass thresholds
in the QCD running coupling constant $\alpha_s$, based on the step
approximation to the $\beta$--function, is investigated. The errors of
extrapolating $\ase(\mu)$ from low energies to $\mu=M_Z$, implied by
this approximation, are shown to be of the same magnitude as the typical
next--to--next--to--leading order contributions to \as and tend to
increase the resulting \as. The importance of the proper choice of
matching points is emphasized.

\vspace*{1cm}
\noindent
In this note I shall discuss the importance of the proper
treatment of quark mass effects on the QCD running coupling constant
\as. Although the quark mass effects are not large, steady
improvement in the precision of experimental data, coming in particular
from a new generation of experiments at CERN and Fermilab, combined with
significant progress in higher order QCD calculations, have led
to a renewed interest in quantitative aspects of these effects
\cite{Bernreut,Schirkov}. There are essentially two reasons for it.

The first concerns the exploitation of the
next--to--next--to--leading--order (NNLO) QCD calculation that have
recently become available for a number of physical quantities
\cite{Gor1,Gor2,Lar1,Lar2} and which exist basically for massless quarks
only. The NNLO corrections are tiny effects and to include
them makes sense only if they are large compared to errors resulting
from the approximate treatment of the quark mass thresholds in
massless QCD. To quantify the importance of the NNLO
correction to the couplant $\alpha_s$,
consider the difference between the
values of $\ase(\bbare{MS},M_Z)$ in the NLO and NNLO approximations,
taking, as an example, $\Lambda^{(5)}_{\bbare{MS}}=0.2$ GeV. At the NLO
the couplant $a\equiv \ase/\pi$ is given as a solution to the equation
\be
b\ln\frac{\mu}{\Lambda^{(5)}}=\frac{1}{a}+c\ln\frac{ca}{1+ca}
\label{NLO}
\ee
where for $n_f$ massless quarks
\be
b=\frac{33-2n_{f}}{6};\;\;\;c=\frac{153-19n_{f}}{33-4n_f}
\label{bc}
\ee
are the first two, universal, coefficients of the QCD
$\beta$--function appearing on the r.h.s. of the definition equation for
the couplant $a(\mu)$
\be
\frac{\mr{d}a(\mu)}{\mr{d}\ln\mu}\equiv \beta(a)=
-ba^2(\mu)\left(1+ca(\mu)+c_{2}a^2(\mu)+\cdots\right).
\label{RG}
\ee
For $\mu=M_Z$, $n_f=5$ and $\Lambda^{(5)}_{\bbare{MS}}=0.2$ GeV we get
$a^{\mr{NLO}}(\bbare{MS},M_Z)=0.03742$. At the NNLO and for
$c_2>c^2/4$ we have instead of \reff{NLO}
\be
b\ln\frac{\mu}{\Lambda} =  \frac{1}{a}+c\ln\frac{ca}
{\sqrt{1+ca+c_2a^2}}+f(a,c_2)
\label{NNLO}
\ee
\be
f(a,c_2)= \frac{2c_2 -c^2}{d}\left(\arctan\frac{2c_2a+c}{d}
-\arctan\frac{c}{d}\right);\;\;\;d\equiv \sqrt{4c_2-c^2}
\label{f}
\ee
where $c_2(\bbare{MS},n_f=5)=1.475$.
Combining \reff{NNLO} and \reff{f} yields $a^{\mr{NNLO}}
(\bbare{MS},M_Z)=0.03665$. The relative difference between these two
approximations
\be
\frac{a^{\mr{NLO}}-a^{\mr{NNLO}}}{a^{\mr{NLO}}}\doteq 0.02
\label{relcon}
\ee
thus amounts to about 2\%. The NNLO correction should therefore be
included only if the neglected effects can be expected to be
smaller than this number. However, as we shall see, quark mass threshold
effects can in some circumstances be just of this magnitude!
In general the magnitude of higher order corrections to \as depends on
the renormalization scheme employed and so does also the estimate
\reff{relcon}. However, if defined as the relative difference between
the NLO and NNLO approximations this dependence is weak.

The second reason is related to the problem of comparing the values
of $\alpha_s$,
determined from different quantities characterized by vastly
different momentum scales. As recently emphasized in an extensive
review on \as determinations \cite{Bethke}, there is small, but
nonnegligible discrepancy between the
value of $\ase(\bbare{MS},M_Z)$ obtained by extrapolation from some of
the low energy quantities and  $\ase(\bbare{MS},M_Z)$ determined
directly at the scale $M_Z$ at LEP, the latter giving the value higher
by about 5--10\%. Simultaneously, it has been noted in \cite{Bethke}
that there is an exception to this behaviour in the case of the ratio
$R_{\tau}$ \cite{Pich1}, which, when extrapolated from $m_{\tau}$ to
$M_Z$, gives values of \as close to that measured directly at LEP
($0.120\pm0.005$ \cite{Pich2}) as well. The physical relevance of the
discrepancy between the low energy extrapolations and direct
measurements of \as at LEP has very recently been emphasized by Schifman
in \cite{Schifman}. In particular, he argues that the extrapolations
of $\ase(m_{\tau})$ to $\ase(M_Z)$ are unreliable due to problems with
the control of power corrections. As will be shown below, a part of this
overestimate of the extrapolated value of $\ase(M_Z)$ may in fact be
simply due to the approximate treatment of the $c$ and $b$ quark
thresholds.

In the rest of this note I shall analyze the quantitative consequences
of the exact treatment of quarks mass thresholds at the LO and formulate
the conventional matching procedure \cite{Mar} for massless quarks in
such a way that its results are so close to the exact ones that
the available NNLO calculation can be consistently included.
I shall describe in detail the approximation in which the ``light''
quarks $u,d$ and $s$ are considered massless, while the $c,b$ and $t$
quarks remain massive.

As complete
multiloop calculations with massive quarks are very complicated and
are available only at the leading order, all higher order
phenomenological analyses use the calculations with a fixed effective
number $n_f$ of massless quarks, depending on the characteristic scale
of the quantity. In order to relate two regions of different effective
numbers of massless quarks the approximate matching procedure developed
in \cite{Mar} is commonly used. It should be emphasized that this
procedure concerns only those mass effects that can be absorbed in the
renormalized couplant.  At higher orders there are, however, mass
effects that remain in the expansion coefficients even after the effects
of heavy quarks have been absorbed in a suitably defined running
couplant.

In QCD with (at least some) massive quarks the renormalization group
equation for the couplant $a(\mu)$ looks formally as in massless QCD.
The only, but important, difference concerns the two lowest order
$\beta$--function coefficients, $b,bc$, which are no longer
unique as in massless QCD, but may depend on the scale $\mu$.
While in the class of \bbar{MS}--like renormalization conventions
$b=11/2-n_f/3$ as in massless QCD, in MOM--like ones it becomes a
nontrivial function of the scale $\mu$ \cite{RG}
\be
b(\mu/m_i)=\frac{11}{2}-\frac{1}{3}\sum_{i} h_i(x_i),\;\;\;\;
x_i\equiv\frac{\mu}{m_i}
\label{xi}
\ee
where the sum runs over all the quarks considered, $m_i$ are the
corresponding renormalized quark masses
\footnote{For the purpose of this discussion quark masses are regarded
as constants, altough they too get renormalized and thus ``run''
as well.}
and the threshold function $h(x)$ is given as \cite{RG}
\be
h(x)\equiv 6x^2 \int_0^1 \mr{d}z \frac{z^2(1-z)^2}{1+x^2 z(1-z)}=
1-\frac{6}{x^2}+\frac{12}{x^3\sqrt{4+x^2}}
\ln\frac{\sqrt{4+x^2}+x}{\sqrt{4+x^2}-x}
\doteq \frac{x^2}{5+x^2}.
\label{h(x)}
\ee
The last - approximate - equality is a very accurate
approximation to the exact form of $h(x)$ in the whole range
$x\in (0,\infty)$ \cite{RG}. This allows a simple treatment of the
quark mass thresholds at the LO.  There is no analogous calculation
of the
next $\beta$--function coefficient, $c$, for massive quarks and thus
no possibility to repeat the present analysis at the NLO. This is one
of the reasons why most of the phenomenological analyses use the so
called ``step'' approximation in which at any value of $\mu$ one works
with a finite effective number of massless quarks, which changes
discontinuously at some matching points $\mu_i$. Consequently $b(n_f)$
becomes effectively a function of $\mu$, discontinuous at these
matching points, as shown in the Fig.1a.

\begin{figure}
\begin{center}
\epsfig{file=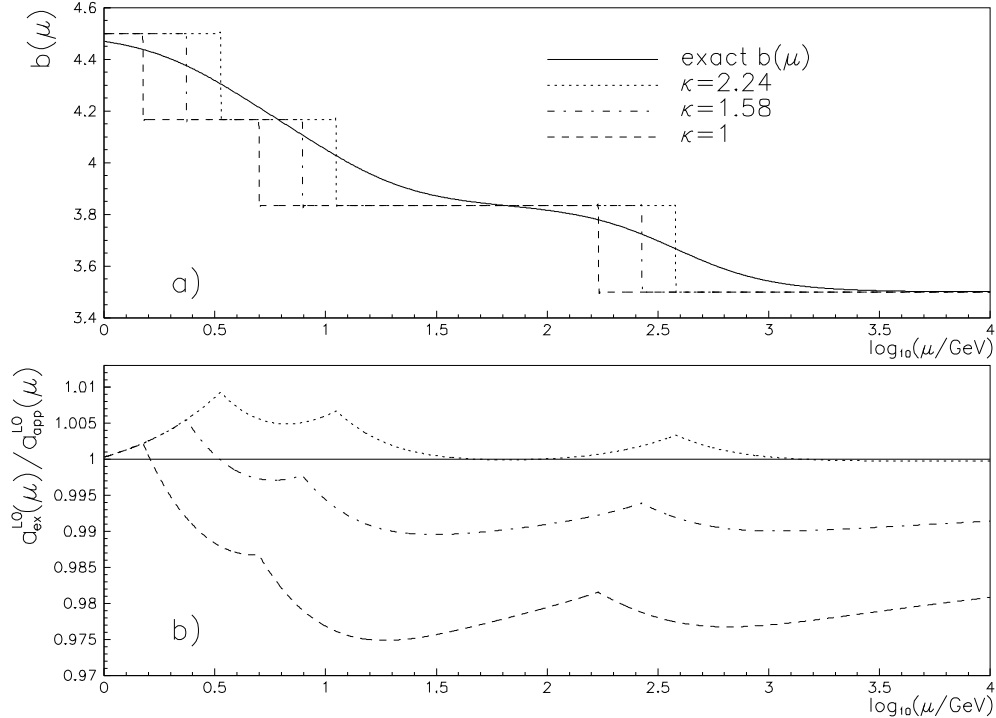,height=8cm}
\end{center}
\caption{a) $b(\mu/m_i)$ together with three step
approximations, corresponding to matching at the points
$\mu_{i}=\kappa m_i;i=c,b,t$ with $\kappa=1,1.58,2.24$;
b) the ratio $R_a$ for the same three approximations, made to
coincide at $\mu_0=1$ GeV.}
\end{figure}

The matching points are assumed proportional to the masses of the
corresponding quarks, $\mu_i\equiv \kappa m_i$. In principle different
quark threshold could be associated with different $\kappa$, but for
simplicity I take them equal. The
free parameter $\kappa$, allowing for the variation of the
proportionality factor, turns out to be quite important for the accuracy
of the step approximation. At the LO the matching procedure then
consists in
the following relations at the matching points $\mu_i$ (the number
in the superscript defines the corresponding effective number of
massless quarks)
\begin{eqnarray}
a^{\mr{LO},3}_{app}(\kappa m_{c}/\Lambda^{(3)}) & = &
a^{\mr{LO},4}_{app}(\kappa m_{c}/\Lambda^{(4)})
\Rightarrow \Lambda^{(4)}=\Lambda^{(3)}\left(\frac{\Lambda^{(3)}}
{\kappa m_c}\right)^{1/3b(4)}
\label{mc} \\
a^{\mr{LO},4}_{app}(\kappa m_{b}/\Lambda^{(4)}) & = &
a^{\mr{LO},5}_{app}(\kappa m_{b}/\Lambda^{(5)})
\Rightarrow \Lambda^{(5)}=\Lambda^{(4)}\left(\frac{\Lambda^{(4)}}
{\kappa m_b}\right)^{1/3b(5)}
\label{mb} \\
a^{\mr{LO},5}_{app}(\kappa m_{t}/\Lambda^{(5)}) & = &
a^{\mr{LO},6}_{app}(\kappa m_{t}/\Lambda^{(6)})
\Rightarrow \Lambda^{(6)}=\Lambda^{(5)}\left(\frac{\Lambda^{(5)}}
{\kappa m_t}\right)^{1/3b(6)}
\label{mt}
\end{eqnarray}
Note that each of the intervals of fixed $n_f$
is associated with a different value of the
$\Lambda$--parameter, $\Lambda^{(n_f)}$. The resulting dependence
$a(\mu/m_i)$ on $\mu$ is thus continuous at each of the
matching points, but its derivatives at these points are discontinuous,
reflecting the discontinuity of the step approximations to $b(\mu/m_i)$.
This procedure can be easily extended to any finite order. Let me point
out that the more
sophisticated procedure for matching the couplants
corresponding to different effective $n_f$ developed in \cite{BW}
coincides to the LO with \reff{mc}--\reff{mt}.

To estimate the errors in \as resulting from the
above defined approximate treatment of quark thresholds we merely need
solve the LO equation with exact explicit mass
dependence as given in \reff{xi}
\be
\frac{\mr{d}a(\mu)}{\mr{d}\ln \mu}=-a^2\left(\frac{11}{2}-\frac{1}{3}
\sum_{i=1}^{6} h(x_i)\right).
\label{RGhx}
\ee
For our purposes the approximation $h(x)\doteq x^2/(5+x^2)$
is entirely adequate and yields
\be
a(\mu)=\frac{1}{\ds {\left(\frac{11}{2}-\frac{3}{3}\right) \ln
\frac{\mu}{\Lambda^{(3)}} -\frac{1}{3} \sum_{i=c,b,t}\ln
\frac{\sqrt{\mu^2+5m_i^2}}
{\sqrt{\left(\Lambda^{(3)}\right)^2+5m_i^2}}}}
\label{am}
\ee
where the fraction $\frac{3}{3}$ comes from the sum over the three
massless quarks $u,d,s$ and $\Lambda^{(3)}$ is the corresponding
$\Lambda$--parameter appropriate to 3 massless quarks.
For the heavy quarks $c,b$ and $t$ I take in the following $m_c=1.5$
GeV, $m_b=5$ GeV, $m_t=170$ GeV.  The distinction between the ``light''
and ``heavy'' quarks is given by the relative magnitude of
$m_i$ and $\Lambda$, the latter being defined by the condition
$5m_i^2\gg\Lambda$. For the above values of $m_c,m_b,m_t$ this condition
 is very well satisfied.  Consequently, for $\mu \ll m_i,i=c,b,t$
\reff{am} approaches smoothly $a^{\mr{LO}}$ for $n_f$=3, while for $\mu
\gg m_i$ and neglecting $\Lambda^{(3)}$ with respect to $5m_i^2$,
it goes to
\be
a(\mu)=\frac{1}{\ds b(6)\ln \frac{\mu}{\Lambda^{(3)}}+ \frac{1}{3}
\ln\left( \frac{\sqrt{5}m_c}{\Lambda^{(3)}}
\frac{\sqrt{5}m_b}{\Lambda^{(3)}} \frac{\sqrt{5}m_t}{\Lambda^{(3)}}
\right)}= \frac{1}{\ds b(6)\ln \frac{\mu}{\Lambda^{(6)}
(\sqrt{5})}}
\label{acbt}
\ee
where the parameter $\Lambda^{(6)}(\kappa)$ depends in general on
$\kappa$ and
\be
\Lambda^{(6)}(\sqrt{5})
\equiv \Lambda^{(3)}
\left(\frac{\Lambda^{(3)}}{\sqrt{5}m_c}
\frac{\Lambda^{(3)}}{\sqrt{5}m_b}
\frac{\Lambda^{(3)}}{\sqrt{5}m_t}\right)
^{\frac{1}{3b(6)}}
=\left(\frac{1}{\sqrt{5}}\right)^{\frac{1}{b(6)}}
\Lambda^{(6)}(1)
\label{Lambda6}
\ee
coincides with $\Lambda^{(6)}$ defined via the
subsequent application of the matching
relations \reff{mc}--\reff{mt} for $\kappa=\sqrt{5}\doteq 2.24$.
Although from the point of view of the matching
procedure $\kappa$ is not exactly fixed, the value
$\kappa=\sqrt{5}$ will be shown to be in some sense the best choice.
The relation between $\Lambda^{(6)}$ and $\Lambda^{(3)}$
depends nontrivially on $\kappa$.

In Fig. 1a $b(\mu)$ is plotted as a function of $\mu$ for the above
mentioned masses of $c,b$ and $t$ quarks, together with its step
approximations corresponding to three different values of
$\kappa=1,\sqrt{5},\sqrt{5/2}$.
There is hardly any sign of the steplike behaviour of the function
$h(x)$ in the region of the $c$ and $b$ quark thresholds and only a
very unpronounced indication of the plateau between the $b$ and $t$
quark thresholds. The steplike approximations are poor representations
of the exact $h(x)$ primarily due to the rather slow approach of $h(x)$
to unity as $x\rightarrow \infty$. However, there is a marked
difference between the three approximations. While the step
approximation with the conventional choice $\kappa=1$
understimates the true $h(x)$ in the whole interval displayed and would
do so even when some smoothing were applied, $\kappa=\sqrt{5}$ gives
clearly much better approximation as the corresponding curve is
intersected by the exact $h(x)$  at about the middle of each step.

In Fig. 1b the $\mu$ dependence of the ratio
\be
R_a\equiv\frac{a^{\mr{LO}}_{ex}(\mu)}{a^{\mr{LO}}_{app}(\mu)}
\label{app}
\ee
between the above exact solution \reff{acbt} and the approximate
expressions for the above mentioned values of $\kappa$
is plotted assuming $\Lambda^{(3)}=200$ MeV
\footnote{
The resulting ratio $R_a(M_Z)$ depends, however, only weekly on
$\Lambda^{(3)}$ in the interval $(200,400)$ MeV.}.
As we basicaly want to compare the results of
different extrapolations starting from the same initial $\mu_0$,
$\Lambda^{(3)}$ used in the approximate solutions was rescaled by the
factor 1.004 with respect to $\Lambda^{(3)}$ in \reff{am},
thereby guaranteeing that all expressions coincide at $\mu_0=1$ GeV.
Any deviation from unity in Fig. 1b is then entirely the effect
of an approximate treatment of the heavy quark thresholds. The Fig.1
contains several simple messages.

The approximate solutions based on the matching procedure
defined in \reff{mc}--\reff{mt} are in general much
better immediately {\em below} the matching point than
above it and is worst at about $5m_{\mr{match}}$. This reflects the
fact that the function $h(x)$ vanishes fast (like $x^2$) at zero
but approaches unity only very slowly. Moreover, in the $M_{Z}$ range
the effect of the $c$ quark threshold is essentially the same as
that of the $b$ quark and both are much more important than that of the
top quark, although $M_Z/m_c\approx 60$, $M_{Z}/m_b\approx 18$, while
$M_{Z}/m_t \approx 1/2$!

The effect of varying $\kappa$ is quite important, in particular
with respect to the $c$ and $b$ quark thresholds.
In general $\kappa >1$ improves the
approximation above the matching point, but worsens it immediately below
it. The choice $\kappa=\sqrt{5}$, suggested by the asymptotic behaviour
of \reff{acbt} is clearly superior
practically in the whole displayed interval $\mu\in(1,10^4)$ GeV
and leads to an excellent (on the level of 0.1\%) agreement
with the exact solution in this interval.
On the contrary the conventional choice $\kappa=1$
leads to much larger deviation from the exact result, which exceeds
$2$\% in most of this region. This discrepancy is
thus of the same magnitude
as the effects of NNLO corrections to the couplant itself.
It turns out that the effect of an exact treatment of the quark
mass thresholds is as important as that
of the NNLO correction to the $\beta$--function and must
therefore be taken into account, whenever the latter is considered.

\begin{figure}
\begin{center}
\epsfig{file=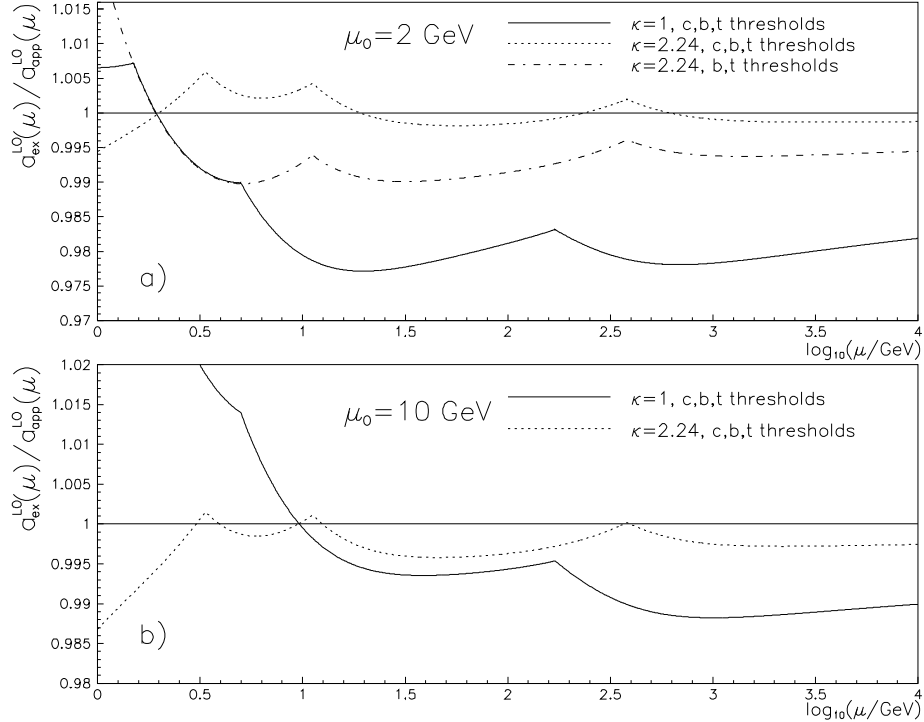,height=8cm}
\end{center}
\caption{The same as in Fig. 1b, but for $\mu_0$ equal to 2 and 10
GeV.}
\end{figure}

In Fig. 1 the exact as well as approximate expressions for the couplant
are made to concide at $\mu_0=1$ GeV, in order to stay below the $c$
quark threshold. In actual extrapolations one usually starts at somewhat
higher scales. For the $\tau$--lepton decay rate as well as the GLS sume
rule, the corresponding initial $\mu_0$ is about 2 GeV, for the various
$\Upsilon$ characteristics $\mu_0\approx 10$ GeV. In Fig. 2 the ratio
\reff{app} is therefore
plotted for these two starting scales and the indicated
values of $\kappa$. We see that
\bit
\item For $\mu_0=2$ GeV the situation is essentially the
same as for $\mu_0=1$ GeV, i.e. extrapolated
$\ase^{\mr{LO}}(M_Z)$ is
about 2\% higher when the step approximation with $\kappa=1$ is used.
The fact that conventional
extrapolations of $\ase(m_\tau)$ to $\ase(M_Z)$ use this
step approximation with $\kappa=1$ could thus cause part of
the overestimate of $\ase(M_Z)$ discussed in \cite{Schifman}.
On the other hand the choice $\kappa=\sqrt{5}$ again leads to an
excellent agreement with the exact extrapolation above $10$ GeV and
very satisfactory even down to 1 GeV.
\item For $\mu_0=10$ GeV the choice $\kappa=\sqrt{5}$ is only marginally
better that $\kappa=1$ when extrapolating to $\mu=M_Z$, but now both
values yield $\ase^{\mr{LO}}(M_Z)$ only about 0.5\% from the
exact LO result. This observation is consistent with the fact that
$\ase(M_Z)$ extrapolated from $\Upsilon$ characteristics is lower than
that from $R_{\tau}$ and consistent with those based on deep inelastic
scattering. For this choice of $\mu_0$
the superiority of the choice $\kappa=\sqrt{5}$ is
obvious when extrapolating back, i.e. to lower scales $\mu$. Indeed,
this is to be expected from Fig. 1b and 2a.
\item If for $\mu_0=2$ GeV and $\kappa=\sqrt{5}$ only the $b$ and $t$
quark thresholds (dash--dotted curve in Fig. 2a) are considered
\footnote{In these circumstances and for $\kappa=\sqrt{5}$ the matching
point corresponding to charmed quark, given by
$\sqrt{5}m_c\doteq 3.26$ is \underline{below} $\mu_0$.},
the result is markedly further away from the the exact
extrapolation than if also the $c$ quark threshold is taken into
account. This illustrates the influence of the
proper treatment of charmed
quark threshold even at the energy scales around $M_Z$.
\eit

The influence
of the choice of $\kappa$ on the results of the matching procedure has
recently been investigated, within the approach introduced in \cite{BW},
in \cite{Bernreut}. Although this matching procedure is
more sophisticated than that based on the relations
\reff{mc}--\reff{mt} and is applicable at any order, it
relates couplants corresponding to different numbers of massless quarks
only and thus does not concern the difference between the results
using the exact expression for the function $h(x)$ and those based on
its step approximations, as investigated in this note. The conclusion
reached there, i.e. that $\ase(M_Z)$ obtained by extrapolation from low
energy scales changes little (by less than 0.6\%) when the matching
points are varied in some reasonable interval, is therefore not in
contradiction with the result of this note.

In summary, the approximate treatment of quark mass thresholds, based
on the step--like approximation of the QCD $\beta$--function and using
the matching procedure of \cite{Mar} with $\kappa=1$, leads to an
overestimate of $\ase^{\mr{LO}}(M_Z)$ extrapolated from low energy
quantities, in particular the $R_{\tau}$ ratio, by about 2\%.
Though not a large effect, it is of the same
magnitude as the typical NNLO corrections to
$\ase^{\mr{NLO}}(\bbare{MS},M_Z)$. This overestimate of the
extrapolated $\ase(M_Z)$ may explain part of the discrepancy between the
extrapolations based on $R_{\tau}$ and deep inelastic scattering.
 On the other hand, the choice of $\kappa=\sqrt{5}$, suggested by
theoretical considerations, yields extrapolations which are in much
better agreement with those based on the exact form of the threshold
function $h(x)$. The analysis of quark mass effects, presented above,
holds strictly speaking, at the LO only. Nevertheless as both the mass
effects the higher order perturbative corrections are small effects,
it seems reasonable to expect that the main conclusions of the present
analysis have more general validity.

\vspace*{0.5cm}
\noindent
I am gratefull to P. Kol\'{a}\v{r} and J. Rame\v{s} for carefull
reading of the manuscript and useful comments and suggestions.

\end{document}